\pdfoutput=1
\documentclass[10pt,a4paper,english,journal]{IEEEtran}
\usepackage[T1]{fontenc}
\usepackage[latin9]{inputenc}
\usepackage{array}
\usepackage{multirow}
\usepackage{amsmath}
\usepackage{amssymb}
\usepackage{stackrel}
\usepackage{graphicx}

\makeatletter

\pdfpageheight\paperheight
\pdfpagewidth\paperwidth

\providecommand{\tabularnewline}{\\}

\usepackage{xcolor,soul}
\sethlcolor{lightgray}

\makeatother

\usepackage{babel}
\begin{document}

\title{Quasi-Dynamic  Frame Coordination For Ultra- Reliability and Low-Latency
in 5G TDD Systems }

\author{\IEEEauthorblockN{Ali A. Esswie$^{1,2}$,\textit{ }Klaus I. Pedersen\textit{$^{1,2}$,}
and\textit{ }Preben E. Mogensen\textit{$^{1,2}$}\\
$^{1}$Nokia Bell-Labs, Aalborg, Denmark\\
$^{2}$Department of Electronic Systems, Aalborg University, Denmark}}

\maketitle
$\pagenumbering{gobble}$
\begin{abstract}
The fifth generation (5G) mobile technology features the ultra-reliable
and low-latency communications (URLLC) as a major service class. URLLC
applications demand a tight radio latency with extreme link reliability.
In 5G dynamic time division duplexing (TDD) systems, URLLC requirements
become further challenging to achieve due to the severe and fast-varying
cross link interference (CLI) and the switching time of the radio
frame configurations (RFCs). In this work, we propose a quasi-dynamic
inter-cell frame coordination algorithm using hybrid frame design
and a cyclic-offset-based RFC code-book. The proposed solution adaptively
updates the RFCs in time such that both the average CLI and the user-centric
radio latency are minimized. Compared to state-of-the-art dynamic
TDD studies, the proposed scheme shows a significant improvement in
the URLLC outage latency, i.e., $\sim92\%$ reduction gain, while
boosting the cell-edge capacity by $\sim189\%$ and with a greatly
reduced coordination overhead space, limited to $\textnormal{B-bit}.$ 

\end{abstract}

\section{Introduction}

Ultra-reliable low-latency communication (URLLC) is a key driver of
the fifth generation (5G) mobile networks {[}1{]}. Various URLLC use
cases require one-way radio latency of one or several milliseconds
with an outage probability below $10^{-5}$ {[}2{]}. As most of the
5G URLLC deployments are envisioned over the 3.5 GHz band, the time
division duplexing (TDD) becomes a vital candidate transmission mode
due to its frame adaptation, in order to dynamically match the sporadic
URLLC capacity in both downlink (DL) and uplink (UL) directions {[}3{]}. 

With the 5G new radio (NR), the agile frame structure with variable
transmission time interval (TTI) duration is introduced {[}3, 4{]}.
Thus, 5G-NR TDD offers more adaptation flexibility with much faster
link-direction update periodicity, that is slot-dependent instead
of being frame-based, i.e., $\leq1$ ms. However, the coexistence
of different transmission directions in adjacent cells results in
cross link interference (CLI) {[}5{]}, i.e., base-station to base-station
(BS-BS) and user-equipment to user-equipment (UE-UE) CLI, respectively.
Hence, URLLC performance is highly impacted by the degraded decoding
ability, due to the fast-varying CLI, and the waiting interval to
the first DL/UL transmission opportunity. 

To the best of our knowledge, no prior work has assessed the performance
of the URLLC outage with the 5G-NR dynamic TDD technology. The state-of-the-art
TDD proposals consider joint multi-cell scheduling, cell muting, and
enhanced power control {[}6, 7{]} to minimize the average network
CLI. Furthermore, advanced massive multi-antenna processing and beam-forming
{[}8{]} are envisioned as vital to counteract the CLI by utilizing
the channel hardening phenomenon. Opportunistic inter-cell coordination
algorithms {[}9{]} are also proven attractive to boost the cell capacity
of the dynamic TDD systems; however, at the expense of a sub-optimal
URLLC outage performance. 

In this work, we propose a hybrid-frame based coordination scheme
(HFCS) for 5G-NR dynamic TDD systems. The proposed HFCS introduces
a multi-objective and slot-dependent dynamic user scheduling. A hybrid
radio frame structure and sliding radio frame configuration (RFC)
code-book are designed to virtually extend the degrees of freedom
of the TDD dynamicity. Thus, the URLLC users with the worst radio
conditions always guarantee semi-preemptive, i.e., immediate scheduling
over pre-set time slots, and CLI-free transmissions, leading to a
significant reduction of the URLLC tail latency. The proposed coordination
scheme shows a significant enhancement in the URLLC outage performance
as well as maximizing the ergodic capacity, and with a confined coordination
overhead span. 

The performance of the proposed scheme is assessed by realistic system
level simulations, due to the complexity of the 5G-NR and addressed
problem herein. The major functionalities of the physical and media
access control layers of the 5G-NR are incorporated and calibrated
against latest 3GPP assumptions, including UL and DL channel modeling,
hybrid automatic repeat request (HARQ), adaptive modulation and coding
selection (MCS) and dynamic user scheduling. 

This paper is organized as follows. Section II introduces the system
modeling of this work while Section III presents the problem formulation.
Section IV details the proposed solution and Section V discusses the
numerical results of the proposed scheme. Finally, conclusions are
drawn in Section VII. 

\section{System Modeling }

A macro 5G-NR TDD system is considered, with a single cluster of $C$
cells, each with $N_{t}$ antennas. Each cell has an average of $K^{\textnormal{dl}}$
and $K^{\textnormal{ul}}$ uniformly-distributed DL and UL active
UEs, respectively, each with $M_{r}$ antennas. We assume a URLLC
dedicated network where the sporadic FTP3 traffic is adopted with
finite packet sizes of $\textnormal{\ensuremath{\mathit{f}^{dl}}}$
and $\textnormal{\ensuremath{\mathit{f}^{ul}}}$ bits, and Poisson
arrival processes $\textnormal{\ensuremath{\lambda}}^{\textnormal{dl}}$
and $\textnormal{\ensuremath{\lambda}}^{\textnormal{ul}},$ in the
DL and UL directions. Accordingly, the average offered load per cell
in DL direction is: $K^{\textnormal{dl}}\times\textnormal{\ensuremath{\mathit{f}^{dl}}}\times\textnormal{\ensuremath{\lambda}}^{\textnormal{dl}}$
and in UL direction as: $K^{\textnormal{ul}}\times\textnormal{\ensuremath{\mathit{f}^{ul}}}\times\textnormal{\ensuremath{\lambda}}^{\textnormal{ul}}$.

We assume an RFC of 10 sub-frames, each can be DL, UL or a special
sub-frame. UEs are dynamically multiplexed by the orthogonal frequency
division multiple access with 15 kHz sub-carrier spacing. The smallest
scheduling unit is the physical resource block (PRB) of 12 consecutive
sub-carriers. Furthermore, we adopt a user scheduling per a mini-slot
duration of 7-OFDM symbols for faster URLLC transmissions.

Furthermore, an arbitrary master cell is initially identified in each
cluster, where other cells are considered as slaves. All slave cells
within the cluster are bidirectionally inter-connected to the master
cell through the \textit{Xn interface}.

We define $\text{\ensuremath{\mathfrak{B}}}_{\textnormal{dl}},$ $\text{\ensuremath{\mathfrak{B}}}_{\textnormal{ul}}$,
$\text{\ensuremath{\mathcal{K}}}_{\textnormal{dl}}$ and $\text{\ensuremath{\mathcal{K}}}_{\textnormal{ul}}$
as the inclusive sets of cells and UE with DL and UL transmission
directions, respectively. Hence, the pre-decoding received signal
at the $k^{th}$ UE, where $k\text{\ensuremath{\in\text{\ensuremath{\mathcal{K}}}_{\textnormal{dl}}}}$,
$c_{k}\text{\ensuremath{\in\text{\ensuremath{\mathfrak{B}}}_{\textnormal{dl}}}}$,
is expressed by

{\small{}
\begin{equation}
\boldsymbol{\textnormal{y}}_{k,c_{k}}^{\textnormal{dl}}=\underbrace{\boldsymbol{\textnormal{\textbf{H}}}_{k,c_{k}}^{\textnormal{dl}}\boldsymbol{\textnormal{\textbf{v}}}_{k}s_{k}}_{\text{Useful signal}}+\underbrace{\sum_{i\in\text{\ensuremath{\mathcal{K}}}_{\textnormal{dl}}\backslash k}\boldsymbol{\textnormal{\textbf{H}}}_{k,c_{i}}^{\textnormal{dl}}\boldsymbol{\textnormal{\textbf{v}}}_{i}s_{i}}_{\text{BS to UE interference}}+\underbrace{\sum_{j\in\text{\ensuremath{\mathcal{K}}}_{\textnormal{ul}}}\boldsymbol{\textnormal{\textbf{G}}}_{k,j}\boldsymbol{\textnormal{\textbf{w}}}_{j}s_{j}}_{\text{UE to UE interference}}+\boldsymbol{\textnormal{\textbf{n}}}_{k}^{\textnormal{dl}},
\end{equation}
}where $\boldsymbol{\textnormal{\textbf{H}}}_{k,c_{i}}^{\textnormal{dl}}\in\text{\ensuremath{\mathcal{C}}}^{M_{r}\times N_{t}}$
is the DL fading channel from the cell serving the $i^{th}$ UE, to
the $k^{th}$ UE, $\boldsymbol{\textnormal{\textbf{v}}}_{k}\in\text{\ensuremath{\mathcal{C}}}^{N_{t}\times1}$
, $\boldsymbol{\textnormal{\textbf{w}}}_{j}\in\text{\ensuremath{\mathcal{C}}}^{M_{r}\times1}$
and $s_{k}$ denote the single-stream zero-forcing precoding vector
at the $c_{k}^{th}$ cell, precoding vector at the $j^{th}$ UE, and
transmitted data symbol of the  $k^{th}$ UE, respectively, $\boldsymbol{\textnormal{\textbf{G}}}_{k,j}\in\text{\ensuremath{\mathcal{C}}}^{M_{r}\times M_{r}}$
represents the the cross-link channel between the $k^{th}$ and $j^{th}$
UEs. $\boldsymbol{\textnormal{\textbf{n}}}_{k}^{\textnormal{dl}}$
denotes the additive white Gaussian noise at the $k^{th}$ UE. In
the UL direction, the received signal at the $c_{k}^{th}$ cell, where
$c_{k}\text{\ensuremath{\in\text{\ensuremath{\mathfrak{B}}}_{\textnormal{ul}}}}$
from $k\text{\ensuremath{\in\text{\ensuremath{\mathcal{K}}}_{\textnormal{ul}}}},$
is modeled by

{\small{}
\begin{equation}
\boldsymbol{\textnormal{y}}_{c_{k},k}^{\textnormal{ul}}=\underbrace{\boldsymbol{\textnormal{\textbf{H}}}_{c_{k},k}^{\textnormal{ul}}\boldsymbol{\textnormal{\textbf{w}}}_{k}s_{k}}_{\text{Useful signal}}+\underbrace{\sum_{j\in\text{\ensuremath{\mathcal{K}}}_{\textnormal{ul}}\backslash k}\boldsymbol{\textnormal{\textbf{H}}}_{c_{k},j}^{\textnormal{ul}}\boldsymbol{\textnormal{\textbf{w}}}_{j}s_{j}}_{\text{UE to BS interference}}+\underbrace{\sum_{i\in\text{\ensuremath{\mathcal{K}}}_{\textnormal{dl}}}\boldsymbol{\textnormal{\textbf{Q}}}_{c_{k},c_{i}}\boldsymbol{\textnormal{\textbf{v}}}_{i}s_{i}}_{\text{BS to BS interference}}+\boldsymbol{\textnormal{\textbf{n}}}_{c_{k}}^{\textnormal{ul}},
\end{equation}
}where $\boldsymbol{\textnormal{\textbf{Q}}}_{c_{k},c_{i}}\in\text{\ensuremath{\mathcal{C}}}^{N_{t}\times N_{t}}$
denotes the cross-link fading channel between the cells that serve
the $k^{th}$ and $i^{th}$ UEs, respectively, $k\text{\ensuremath{\in\text{\ensuremath{\mathcal{K}}}_{\textnormal{ul}}}}$
and $i\in\text{\ensuremath{\mathcal{K}}}_{\textnormal{dl}}$. The
pre-detection signal-to-interference-noise-ratio (SINR) in the DL
direction at the $k^{th}$ UE $\gamma_{k}^{\textnormal{dl}}$ and
in the UL direction at the $c_{k}^{th}$ cell $\gamma_{c_{k}}^{\textnormal{ul}}$,
are given by

{\small{}
\begin{equation}
\gamma_{k}^{\textnormal{dl}}=\frac{p_{c_{k}}^{\textnormal{dl}}\left\Vert \boldsymbol{\textnormal{\textbf{H}}}_{k,c_{k}}^{\textnormal{dl}}\boldsymbol{\textnormal{\textbf{v}}}_{k}\right\Vert ^{2}}{\sigma^{2}+\underset{i\in\text{\ensuremath{\mathcal{K}}}_{\textnormal{dl}}\backslash k}{\sum}p_{c_{i}}^{\textnormal{dl}}\left\Vert \boldsymbol{\textnormal{\textbf{H}}}_{k,c_{i}}^{\textnormal{dl}}\boldsymbol{\textnormal{\textbf{v}}}_{i}\right\Vert ^{2}+\underset{j\in\text{\ensuremath{\mathcal{K}}}_{\textnormal{ul}}}{\sum}p_{j}^{\textnormal{ul}}\left\Vert \boldsymbol{\textnormal{\textbf{G}}}_{k,j}\boldsymbol{\textnormal{\textbf{w}}}_{j}\right\Vert ^{2}},
\end{equation}
}{\small \par}

{\small{}
\begin{equation}
\gamma_{c_{k}}^{\textnormal{ul}}=s\frac{p_{k}^{\textnormal{ul}}\left\Vert \boldsymbol{\textnormal{\textbf{H}}}_{c_{k},k}^{\textnormal{ul}}\boldsymbol{\textnormal{\textbf{w}}}_{k}\right\Vert ^{2}}{\sigma^{2}+\underset{j\in\text{\ensuremath{\mathcal{K}}}_{\textnormal{ul}}\backslash k}{\sum}p_{j}^{\textnormal{ul}}\left\Vert \boldsymbol{\textnormal{\textbf{H}}}_{c_{k},j}^{\textnormal{ul}}\boldsymbol{\textnormal{\textbf{w}}}_{j}\right\Vert ^{2}+\underset{i\in\text{\ensuremath{\mathcal{K}}}_{\textnormal{dl}}}{\sum}p_{c_{i}}^{\textnormal{dl}}\left\Vert \boldsymbol{\textnormal{\textbf{Q}}}_{c_{k},c_{i}}\boldsymbol{\textnormal{\textbf{v}}}_{i}\right\Vert ^{2}},
\end{equation}
}where $p_{c_{k}}^{\textnormal{dl}}$ and $p_{k}^{\textnormal{ul}}$
denote the transmission powers of the $c_{k}^{th}$ cell and the $k^{th}$
UE, respectively. Finally, the received UL/DL signals are decoded
by the linear minimum mean square error interference rejection combining
receiver (LMMSE-IRC) {[}4{]} vector $\boldsymbol{\textnormal{\textbf{a}}}$,
expressed as: $\hat{s}_{k}^{\kappa}=\left(\boldsymbol{\textnormal{\textbf{a}}}_{k}^{\kappa}\right)^{\textnormal{H}}\boldsymbol{\textnormal{y}}_{k}^{\kappa},$
$\text{\ensuremath{\mathcal{X}}}^{\kappa},\kappa\text{\ensuremath{\in}}\{\textnormal{ul},\textnormal{dl}\}$,
with $\left(\bullet\right)^{\textnormal{H}}$ as the Hermitian operation.

\section{Problem Formulation }

The URLLC latency and reliability requirements are further challenging
to achieve in 5G-NR dynamic TDD systems, mainly due to the link-direction
switching time and the degraded URLLC decoding performance. The former
is significantly minimized by the flexible 5G frame structure; however,
the latter still remains an open issue.

In fully dynamic TDD macro networks, neighboring cells may have simultaneous
cross-directional transmissions, leading to a strong CLI which varies
per the link-direction update periodicity. With the 5G-NR, such periodicity
is slot-based, i.e., $\leq1$ ms, leading to highly varying CLI fluctuations.
As a result, URLLC UEs inflict significantly degraded decoding performance.
In particular, lower-power URLLC UL transmissions suffer from a strong
CLI from adjacent higher-power DL transmissions, leading to several
HARQ re-transmissions prior to a successful decoding, not satisfying
the URLLC targets. 

Let $u_{c}$ and $d_{c}$ present the estimated numbers of UL and
DL slots during a given RFC while $u_{c}^{\textnormal{opt.}}$ and
$d_{c}^{\textnormal{opt.}}$ are the respective optimal numbers. Hence,
the proposed HFCS defines a programming optimization problem as:
\begin{center}
$R\triangleq\underset{c}{\arg\max}\stackrel[c=1]{C}{\sum}\min\left(\textnormal{\ensuremath{u_{c}},\ensuremath{u_{c}^{\textnormal{opt.}}}}\right)\digamma_{c}^{u}+\min\left(\textnormal{\ensuremath{d_{c}},\ensuremath{d_{c}^{\textnormal{opt.}}}}\right)\digamma_{c}^{d},$
\par\end{center}

subject to:

\begin{equation}
\left\{ \begin{array}{c}
\underset{c}{\arg\min}\,\phi_{c}\left(\eta_{c}\right)=\frac{1}{C}\stackrel[x=1,x\neq c]{C}{\sum}\varphi_{c,x}\left(\eta_{c},\eta_{x}\right),\\
\text{\ensuremath{\forall}}k\in\text{\ensuremath{\mathcal{K}}}_{\textnormal{ul}/\textnormal{dl}}:\underset{k}{\arg\min}\left(\varPsi_{c,k}\right),\,\varPsi_{c,k}\leq\epsilon\,\textnormal{ms},
\end{array}\right.
\end{equation}
where $R$ is total capacity of each cluster, $\digamma_{c}^{u}$
and $\digamma_{c}^{d}$ denote rate utility functions of the UL and
DL transmissions, i.e., capacity gain due to an UL or DL transmission.
$\phi_{c}\left(\eta_{c}\right)$ and $\varphi_{c,x}\left(\eta_{c},\eta_{x}\right)$
represent the average and actual slot misalignment of the requested
RFC by the $c^{th}$ cell $\eta_{c}$ and between the RFCs of the
$c^{th}$ and $x^{th}$ cells, i.e., $\eta_{c}$ and $\eta_{x}$,
respectively, $\forall x\neq c$, and $\varPsi_{c,k}$ is the one-way
radio latency of the $k^{th}$ UL or DL user which is confined by
$\epsilon$ ms. 

For best RFC adaptation and highest ergodic capacity, $u_{c}=u_{c}^{\textnormal{opt.}}$
and $d_{c}=d_{c}^{\textnormal{opt.}}$ should be arbitrarily set in
(5). However, $u_{c}^{\textnormal{opt.}}$ and $d_{c}^{\textnormal{opt.}}$
may introduce a large inter-cell slot misalignment $\phi_{c}$, resulting
in severe CLI within the cluster, and thus, a significant degradation
of the overall capacity $R$ and URLLC latency performance. As such
problem is non-convex, we propose a heuristic approach using complexity-efficient
coordination with hybrid-frame design, multi-objective user scheduling
and a sliding-based RFC code-book. 

\section{Proposed HFCS Coordination }

The proposed HFCS combines a hybrid RFC design, multi-objective distributed
user scheduling, and a cyclic-offset-based RFC code-book. A pre-defined
RFC code-book is constructed and presumed pre-known to all cells within
the cluster, where all RFCs have a set combination of static and dynamic
slots. At each RFC update instant, each slave cell selects the one
RFC from the code-book that most satisfies its individual link-direction
selection criterion. The slave cells signal the index of the selected
RFC to the master cell where it seeks to improve the joint capacity.
Thus, it may slightly change the RFCs requested by slave cells. Accordingly,
the master cell feeds-back the updated RFC indices to the slave cells,
to be adopted until the next RFC update. During each RFC period, each
cell considers a dual-objective dynamic user scheduling.

\subsection{Proposed Inter-Cell Coordination Scheme}

\textbf{\textit{Hybrid RFC design and sliding RFC code-book}}

A hybrid RFC design is adopted, where each RFC is divided into arbitrary
static and dynamic slot sets (SSS, DSS). A SSS denotes the radio slots
which are fixed across all RFCs in the code-book, i.e., static TDD
slots with CLI-free transmissions. However, a DSS implies fully dynamic
radio slots. 

Accordingly, a pre-defined RFC code-book of $\mathcal{N}$ unique
RFCs is constructed such that it is divided into $L$ groups. The
RFCs within each group share the same DL:UL slot ratio, i.e., $d_{c}:u_{c}$;
though, with a different placement during the DSS. For instance, the
DSS of each RFC is a cyclic-shift of the other RFCs, as depicted in
Fig. 1. The structure of the SSS, DSS, and size of the RFC code-book
are design parameters. 

\begin{figure}
\begin{centering}
\includegraphics[scale=0.37]{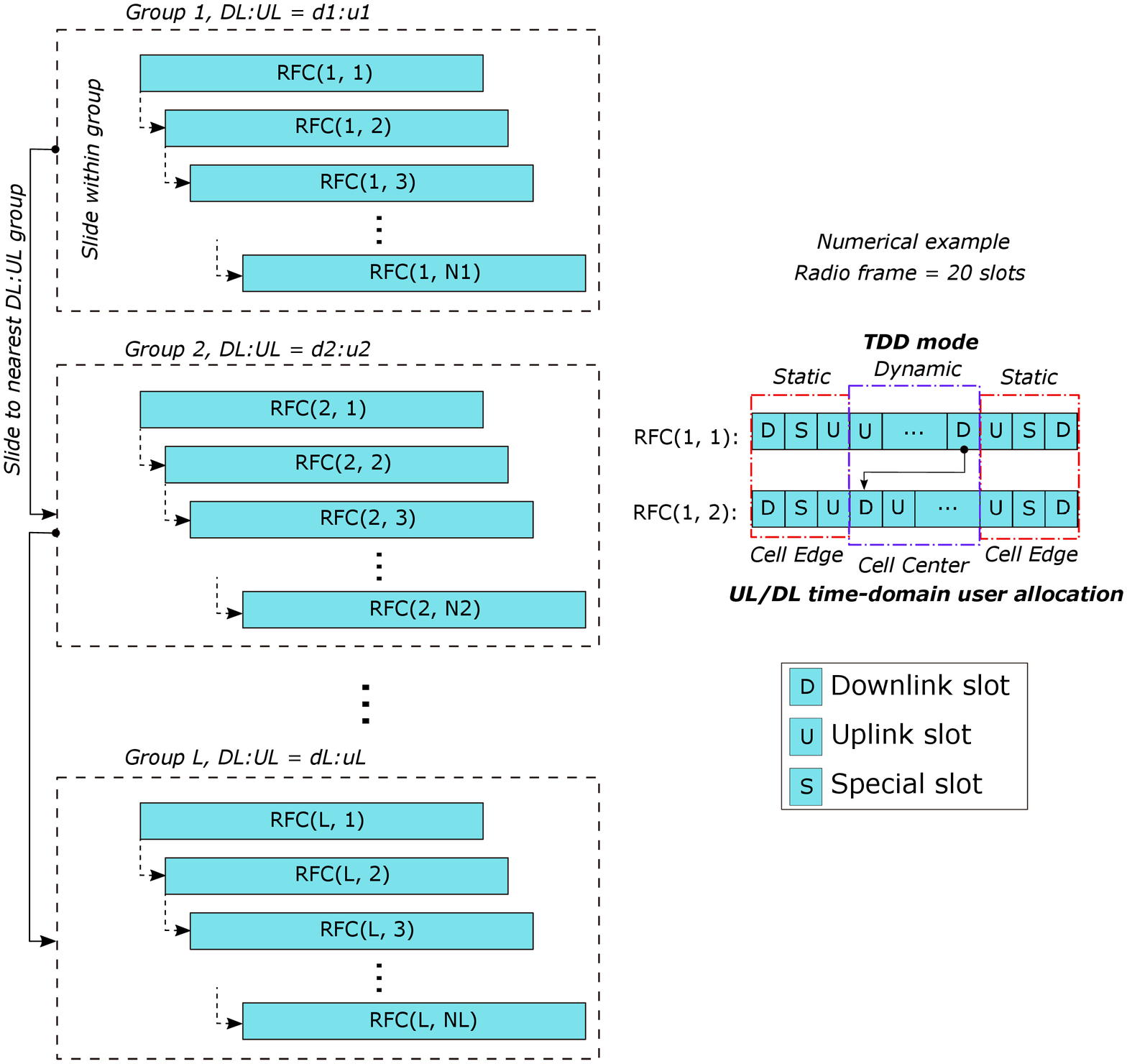}
\par\end{centering}
\centering{}{\small{}Fig. 1. Hybrid RFC design and sliding code-book.}{\small \par}
\end{figure}

\textbf{\textit{At slave cells \textendash{} Traffic and latency adaptation}}

During each RFC update instant, each slave cell selects the one RFC
from the code-book which best satisfies its link-direction selection
criterion. Without loss of generality, we consider the DL/UL buffered
traffic size including pending HARQ re-transmissions as the major
criterion to select the RFC, and accordingly the best $d_{c}:u_{c}$
ratio. Then, the traffic load threshold $\beta_{c}$ is defined as

\begin{equation}
\beta_{c}\leq\frac{\sum Z_{c}^{\textnormal{dl}}}{\sum Z_{c}^{\textnormal{dl}}+\sum Z_{c}^{\textnormal{ul}}},
\end{equation}
where $\sum Z_{c}^{\textnormal{dl}}$ and $\sum Z_{c}^{\textnormal{ul}}$
imply the aggregate traffic in the DL and UL directions, respectively.
With $\beta_{c}=0.5$, if $\sum Z_{c}^{\textnormal{dl}}\gg\sum Z_{c}^{\textnormal{ul}}$,
a cell selects an RFC with a majority of DSS DL slots. Although, the
free selection of the best RFCs requested by each slave cell may result
in severe CLI, hence, several HARQ re-transmissions may be inflicted,
leading to a significant radio latency and reliability. On the other
side, an abrupt change of these RFCs to reduce the average CLI leads
to significant queuing delays up to the first DL/UL transmission opportunities.
Thus, to address the constraints in (5), each cell adaptively estimates
a dynamic sliding threshold $\psi_{c}(t)$, where $t$ is the link-direction
update time, with which it instructs the master cell about the maximum
allowable change of its desired RFC, in order to achieve an adequate
joint URLLC and ergodic capacity performance. 

Let $\Theta^{\textnormal{BS}}$ and $\Theta^{\mathit{\textnormal{UE}}}$
denote the BS-BS and UE-UE CLI at the BS and UE, respectively. These
CLI estimates can be obtained at the BS through radio feedback links
from UEs; however, there is no a standardized mechanism of the CLI
measurement reporting available yet. Then, each BS calculates the
average experienced CLI using an arbitrary filter function. In this
work, we assume a weighted average filter as

\begin{equation}
\Xi_{c}^{\textnormal{avg.}}=\frac{\tilde{\beta}_{c}\times\Theta_{c}^{\textnormal{BS}}+\tilde{\mu_{c}}\times\Theta_{c}^{\mathit{\textnormal{UE}}}}{\Theta_{c}^{\textnormal{BS}}+\Theta_{c}^{\mathit{\textnormal{UE}}}},
\end{equation}

\begin{equation}
\tilde{\beta}_{c},\,\tilde{\mu_{c}}=\left\{ \begin{array}{c}
\frac{1}{\beta_{c}},\frac{1}{\mu_{c}}\,\,\,\forall\,\Theta_{c}^{\textnormal{BS}},\Theta_{c}^{\mathit{\textnormal{UE}}}\leq\varrho\,\textnormal{dBm}\\
\beta_{c},\mu_{c}\,\,\,\forall\,\Theta_{c}^{\textnormal{BS}},\Theta_{c}^{\mathit{\textnormal{UE}}}>\varrho\,\textnormal{dBm}
\end{array}\right.,
\end{equation}

\begin{equation}
\beta_{c}=\frac{\sum Z_{c}^{\textnormal{dl}}}{\sum Z_{c}^{\textnormal{ul}}}\times\mu_{c},
\end{equation}
where $\beta_{c}$ and $\mu_{c}$ are the BS-BS and UE-UE CLI weights,
and $\varrho$ is a CLI threshold. The ratio of both weights is set
to the ratio of the buffered traffic as in (9), such that a cell with
$\sum Z_{c}^{\textnormal{dl}}\gg\sum Z_{c}^{\textnormal{ul}}$, and
accordingly a DL-heavy RFC, shall impose severe BS-BS CLI to adjacent
cells. Hence, under this condition, $\Xi_{c}^{\textnormal{avg.}}$
is biasedly maximized and the RFC adaptation is enforced towards the
CLI minimization. In the delay domain, cells measure the head of line
delay (HoLD) within their DL and UL transmission buffers. HoLD indicates
an estimate of the maximum time required to transmit the last packet
in the buffer, based on the expected UL/DL transmission constraints
of the current RFC. Such metric is of a significant importance with
URLLC since a packet can be considered of no use if its latency deadline
is not fulfilled. Hence, to reduce the average HoLD, selected traffic-based
RFCs should be used without a significant change in order to quickly
transmit the data buffers. 

Thus, we propose a simple and dynamic sliding threshold for a best-effort
trade-off between CLI and radio latency. Fig. 2 shows a numerical
example of such approach. A SSS to DSS ratio of 8:12 is assumed. Thus,
the slot misalignment threshold is bounded by the size of the DSS.
Accordingly, the range of the CLI and HoLD values is quantized over
the DSS size. For an arbitrary cell, if the average CLI, experienced
over the previous measurement cycle, is at maximum, e.g., $\Xi_{c}^{\textnormal{avg.}}=-60$
dBm, it implies a tight slot misalignment threshold should be enforced
to promptly reduce such severe CLI over the upcoming RFC period, e.g.,
$\psi_{c}(t)=1$ slot. However, if such cell simultaneously inflicts
a large HoLD, e.g., $\textnormal{HoLD}$ = 52 ms, the slot misalignment
constraint shall be relaxed, e.g., $\psi_{c}(t)=10$ slots, in order
not to allow the master cell to change the RFC of this cell, hence,
having faster transmissions for the respective traffic. Without loss
of generality, we apply a fair averaging of both misalignment thresholds,
i.e., $\psi_{c}(t)=5$ slots.

\begin{figure}
\begin{centering}
\includegraphics[scale=0.38]{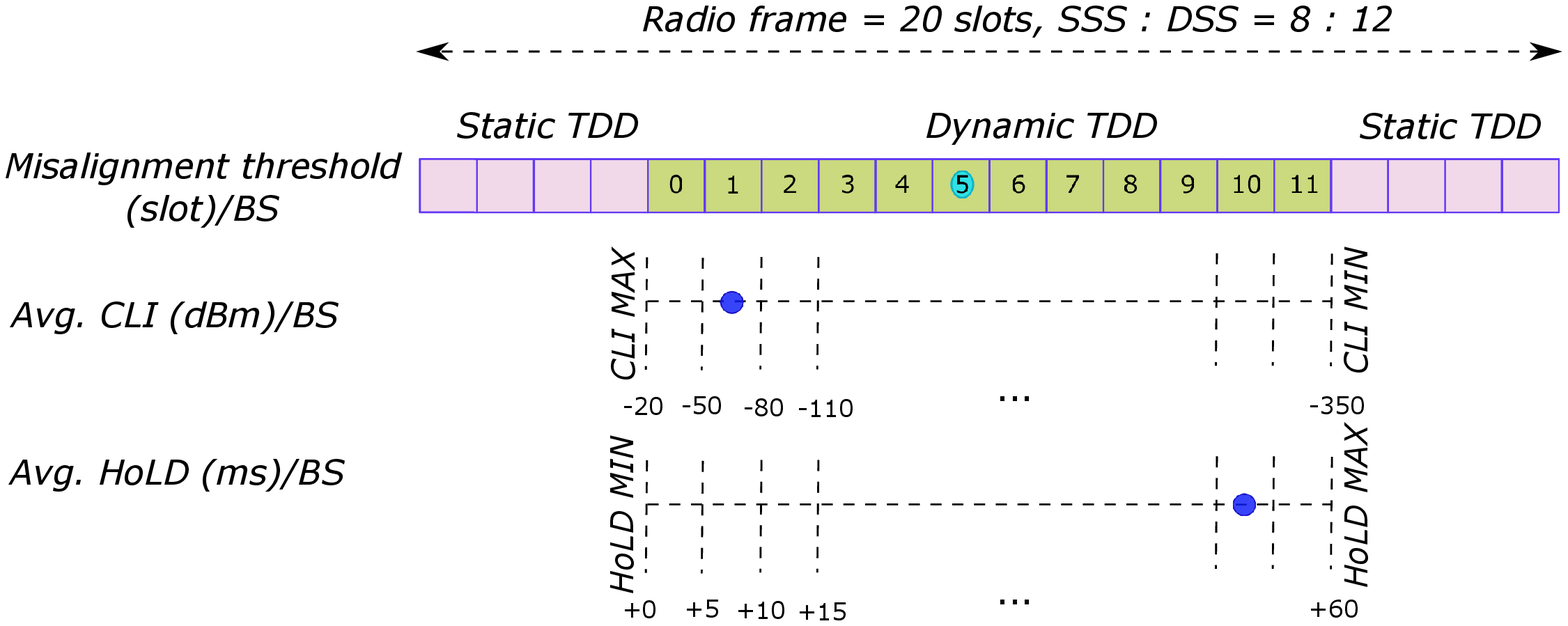}
\par\end{centering}
\centering{}{\small{}Fig. 2. Dynamic misalignment threshold $\psi_{c}(t).$}{\small \par}
\end{figure}

Finally, at each RFC update periodicity, slave cells signal the master
cell with the requested RFC indices of $\textnormal{B}=\log_{2}\left(\text{\ensuremath{\mathcal{N}}}\right)$
bits on the \textit{Xn interface }as well as the maximum allowable
slot misalignment thresholds $\psi_{c}(t).$

\textbf{\textit{At master cell \textendash{} CLI minimization }}

When the master cell receives all RFC information from slave cells,
it first identifies a \textit{common RFC, }which is requested by the
majority of the slave cells. If not feasible, the master cell randomly
selects any reported RFC as the common one, to which all other RFCs
shall maintain the respective slot misalignment thresholds. Thus,
for each RFC $\eta_{c}$ of the $c_{k}^{th}$ cell, master cell calculates
the slot misalignment to the common RFC $\delta_{x},\forall x\neq c$
as in (5). Then, the master cell does not alter such requested RFC
if the following condition is fulfilled:

\begin{equation}
\varphi_{c,x}(t)\leq\psi_{c}(t).
\end{equation}

Hence, the respective slave cell utilizes its best matching RFC to
its latency and capacity outage. Otherwise, the master cell slides
over all RFCs within the same group as the desired RFC of the $c^{th}$
cell $\eta_{c}$. Accordingly, it estimates the corresponding slot
misalignment values and considers the one RFC with $\varphi_{c,x}(t)$
that has the closest linear distance to the requested $\psi_{c}(t).$
If the slot misalignment constraint in (10) is satisfied, master cell
adopts such RFC as the updated RFC of the current cell. This way,
an acceptable average CLI is guaranteed at the slave cells while still
preserving the same requested traffic service ratio $d_{c}:u_{c}$,
leading to a significant improvement of the capacity and outage latency
performance. 

If the slot misalignment constraint is not yet feasible across all
RFCs from the same group as the requested one, master cell progressively
slides to the other RFC groups from the RFC code-book with the nearest
possible $d_{c}:u_{c}$ ratio to the requested ratio, e.g., $d_{c}:u_{c}=4:12\underset{\textnormal{slide\,to}}{\rightarrow}d_{c}^{'}:u_{c}^{'}=3:13$,
and repeats the same process. Herein, the master cell partly relaxes
the target outage requirements of the salve cells due to the abrupt
change in the $d_{c}:u_{c}$ ratio. However, such outage degradation
is bounded across a limited number of slots during the RFC and is
reversely proportional to the size of the RFC code-book $\mathcal{N}$.
As a last best-effort resort, if the constraint in (10) could not
be satisfied across all RFCs, either from same or different group(s),
the master cell considers the one RFC with the closest possible estimated
slot misalignment to desired $\psi_{c}(t)$, and then, it signals
all slave cells within the cluster over the\textit{ Xn interface}
with the updated RFC indices that should be used over the upcoming
RFC periodicity. 

\subsection{Distributed multi-objective user scheduling }

During each RFC periodicity, each cell applies a slot-dependent dynamic
user scheduling. During the DSS instances, cells may adapt an arbitrarily
capacity maximizing user scheduling. Without loss of generality, and
since we assume an equally-prioritized URLLC setup, we adopt the proportional
fair (PF) criterion $\omega$ in both the time and frequency domains
to maintain a global scheduling fairness as

{\small{}
\begin{equation}
\omega\left\{ \textnormal{PF}_{k_{\textnormal{ul/dl}}}\right\} =\frac{r_{_{k_{\textnormal{ul/dl}},rb}}}{\overline{r}_{k_{\textnormal{ul/dl}},rb}},
\end{equation}
}{\small \par}

{\small{}
\begin{equation}
k_{\textnormal{ul/dl}}^{*}=\underset{k_{\textnormal{ul/dl}}\in\text{\ensuremath{\mathcal{K}}}_{\textnormal{ul/dl}}}{\arg\max}\,\,\,\,\omega\left\{ \textnormal{PF}_{k_{\textnormal{ul/dl}}}\right\} ,
\end{equation}
}where $r_{_{k_{\textnormal{ul/dl}},rb}}$ and $\overline{r}_{k_{\textnormal{ul/dl}},rb}$
denote the instantaneous and average delivered rates of the $k^{th}$
UL/DL user. However, during the SSS periods, each cell preemptively
interrupts its individual time-domain scheduling metric by immediately
allocating the users with the worst radio conditions, i.e., potentially
cell-edge users. These users are identified based on the reported
channel quality indication (CQI) reports. To avoid threshold-based
user identification, the UL/DL time-domain scheduler sorts active
users in an ascending-order list in terms of their reported CQI levels,
i.e., users from the top of the list are of worst radio conditions,
thus, scheduler grants them a higher priority for immediate scheduling
during the CLI-free SSS. In the frequency domain, the PF metric is
used to preserve fairness among cell-edge URLLC users. Thus, cell-edge
URLLC users achieve a better decoding ability with faster transmissions,
avoiding the latency-costly HARQ re-transmissions. 

\subsection{Comparison to the state-of-the-art TDD studies}

We evaluate the performance of the proposed scheme against the state-of-the-art
coordinated TDD proposals as:

\textbf{Non-coordinated TDD }(\textbf{NC-TDD}): no RFC coordination
is assumed. Cells independently and dynamically in time pick the RFCs
from the code-book which most meet their individual traffic demand,
as in (6). Hence, maximum TDD RFC flexibility is achieved with no
coordination overhead; however, associated with potentially a large
slot misalignment and severe average CLI levels accordingly. 

\textbf{Sliding code-book based coordinated TDD} (\textbf{SCC-TDD})
{[}9{]}: in our prior work, we introduced a simple inter-cell coordination
algorithm, mainly for broadband services, to significantly reduce
the average slot misalignment, based on a preset global misalignment
threshold $\varOmega$, and hence, the aggregate CLI, resulting in
greatly improved ergodic capacity. Though, it has been demonstrated
not suitable for URLLC transmissions due to the monotonic scheduling
objective. 

\textbf{CLI-free coordinated TDD} (\textbf{CFC-TDD}): cells dynamically
select their respective RFC according to (6). A sophisticated BS-BS
and UE-UE coordination is artificially assumed. That is, BSs and UEs
exchange PRB mapping, UE MCS and precoding information, for them to
perfectly suppress the BS-BS and UE-UE CLI. However, such coordination
introduces a significant control overhead over both the back-haul
and radio interfaces, respectively. In {[}10{]}, a 3GPP technical
study introduces a sub-optimal CFC-TDD approach with a lower overhead
space. However, CFC-TDD holds an optimal theoretical baseline, where
both maximum TDD RFC flexibility and CLI-free transmissions are always
guaranteed. 

\section{Performance Evaluation}

The major simulation assumptions are presented in Table I. During
each TTI, each cell dynamically multiplexes users over system PRBs
using the PF metric, if it is within the DSS of the current RFC or
by preemptive cell-edge user allocations when it is within the SSS.
We consider a fully dynamic MCS selection and adaptive Chase-combining
HARQ re-transmissions, where the HARQ feedback is always prioritized
over new transmissions. The post-detection SINR levels are estimated
by the LMMSE-IRC receiver, where the average interference is identified
by its mean covariance. Finally, we assess the proposed solution under
the latency-efficient user data-gram protocol for several offered
cell loading conditions.

\begin{table}
\caption{{\small{}Simulation parameters.}}
\centering{}%
\begin{tabular}{c|c}
\hline 
Parameter & Value\tabularnewline
\hline 
Environment & 3GPP-UMA, one cluster, 21 cells\tabularnewline
\hline 
UL/DL channel bandwidth & 10 MHz, SCS = 30 KHz, TDD\tabularnewline
\hline 
Antenna setup & $N_{t}=8$ Tx, $M_{r}=2$ Rx\tabularnewline
\hline 
UL power control & LTE-alike, $\alpha=1,\,P0=-103$ dBm\tabularnewline
\hline 
Average user load per cell & $K^{\textnormal{dl}}=K^{\textnormal{ul}}=$ 10 and 20 \tabularnewline
\hline 
TTI configuration & 0.5 ms (7-OFDM symbols)\tabularnewline
\hline 
Traffic model & $\begin{array}{c}
\textnormal{FTP3, \textnormal{\ensuremath{\mathit{f}^{dl}}} = \textnormal{\ensuremath{\mathit{f}^{ul}}} = 400 bits}\\
\textnormal{\ensuremath{\textnormal{\ensuremath{\lambda}}^{\textnormal{dl}}} =167, and 620 pkts/sec}\\
\textnormal{\ensuremath{\textnormal{\ensuremath{\lambda}}^{\textnormal{ul}}} =334, and 620 pkts/sec}
\end{array}$\tabularnewline
\hline 
$\begin{array}{c}
\textnormal{\textnormal{Offered average load per cell}}\\
\textnormal{DL:UL}
\end{array}$ & $\begin{array}{c}
\textnormal{\textnormal{ \,\,\,\,\,\,\,DL:UL = 1:2 (0.6:1.2)} Mbps}\\
\textnormal{\textnormal{DL:UL = 1:1 (5:5)} Mbps}
\end{array}$\tabularnewline
\hline 
Proposed HFCS setup & $\begin{array}{c}
\textnormal{\ensuremath{\mathcal{N}} = 55}\textnormal{ RFCs}\\
L\textnormal{ = 7}\textnormal{ groups}\\
\textnormal{\textnormal{ B = 6 bits}}
\end{array}$\tabularnewline
\hline 
\end{tabular}
\end{table}

Fig. 3 depicts a comparison of the complementary cumulative distribution
function (CCDF) of the URLLC outage latency in the UL direction for
all TDD coordination schemes under assessment, for an average offered
load of 2 Mbps/cell with a DL:UL traffic ratio of 1:2. Furthermore,
we present the latency performance of the best static-TDD case where
the static pattern is pre-selected to perfectly match the DL-to-UL
average traffic ratio, i.e., 6 DL mini-slots, 12 UL mini-slots and
2 guard mini-slots. The optimal CFC-TDD achieves the best URLLC latency
performance, i.e., 42 ms at $10^{-5}$ outage probability. However,
it comes under the ideal assumption of perfect elimination of any
experienced CLI, and with an infinite coordination overhead, which
is infeasible in practice. The proposed HFCS clearly provides a significant
improvement of the UL URLLC latency, approaching the optimal CFC-TDD;
however, with greatly reduced overhead span, mainly limited to $\log_{2}\left(\text{\ensuremath{\mathcal{N}}}\right)$
bits. That is, it achieves $92\%$ and $67\%$ reduction gain in the
UL outage latency compared to SCC-TDD and NC-TDD. The best static-TDD
case out-performs proposed HFCS scheme, i.e., $9\%$ reduction in
the outage latency, due to the absence of the CLI, approaching CFC-TDD;
though, this comes with the assumption that the static RFC pattern
is pre-defined to perfectly align with the traffic demands. 

The significant latency improvements of the proposed HFCS are attributed
to the guaranteed preemptive cell-edge user scheduling with CLI-free
transmissions, where these users majorly control the latency tail,
i.e., outage, performance. Thus, less costly HARQ re-transmissions
are experienced. The SCC-TDD latency performance depends on the preset
misalignment threshold $\varOmega.$ For instance, with a tight $\varOmega=3$,
the master cell may aggressively change the requested RFC of a given
slave cell, in order to only allow for an average misalignment of
three slots. As a result, slave cells may adopt RFCs that do not best
match their current traffic demands, leading to a more queuing delay
to the first transmission opportunity. Finally, the NC-TDD offers
a fair URLLC latency performance since the maximum possible TDD RFC
flexibility is utilized; however, with severe CLI levels.

\begin{figure}
\begin{centering}
\includegraphics[scale=0.6]{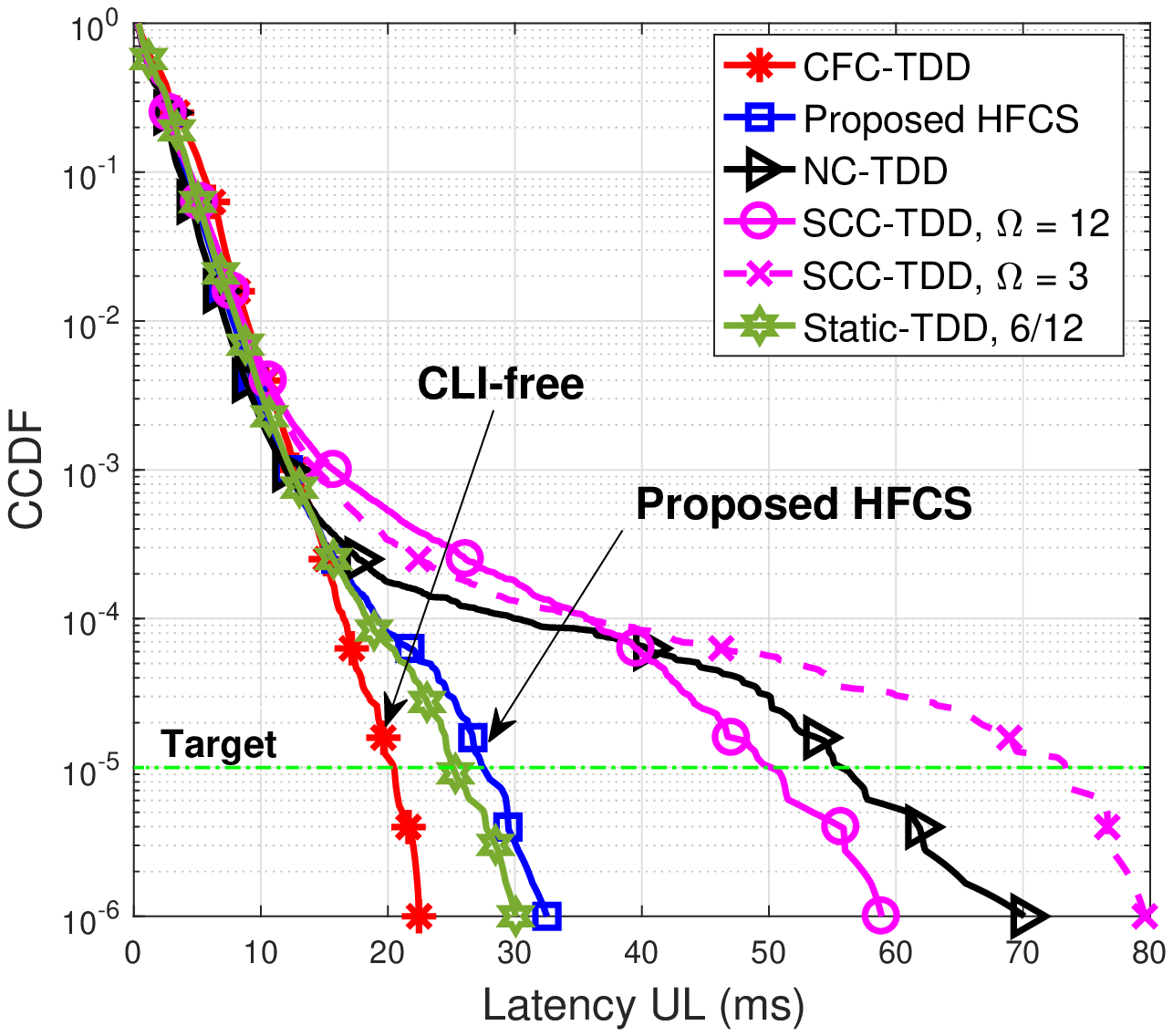}
\par\end{centering}
\centering{}{\small{}Fig. 3. URLLC outage latency in UL direction
(ms).}{\small \par}
\end{figure}
 
\begin{figure}
\begin{centering}
\includegraphics[scale=0.6]{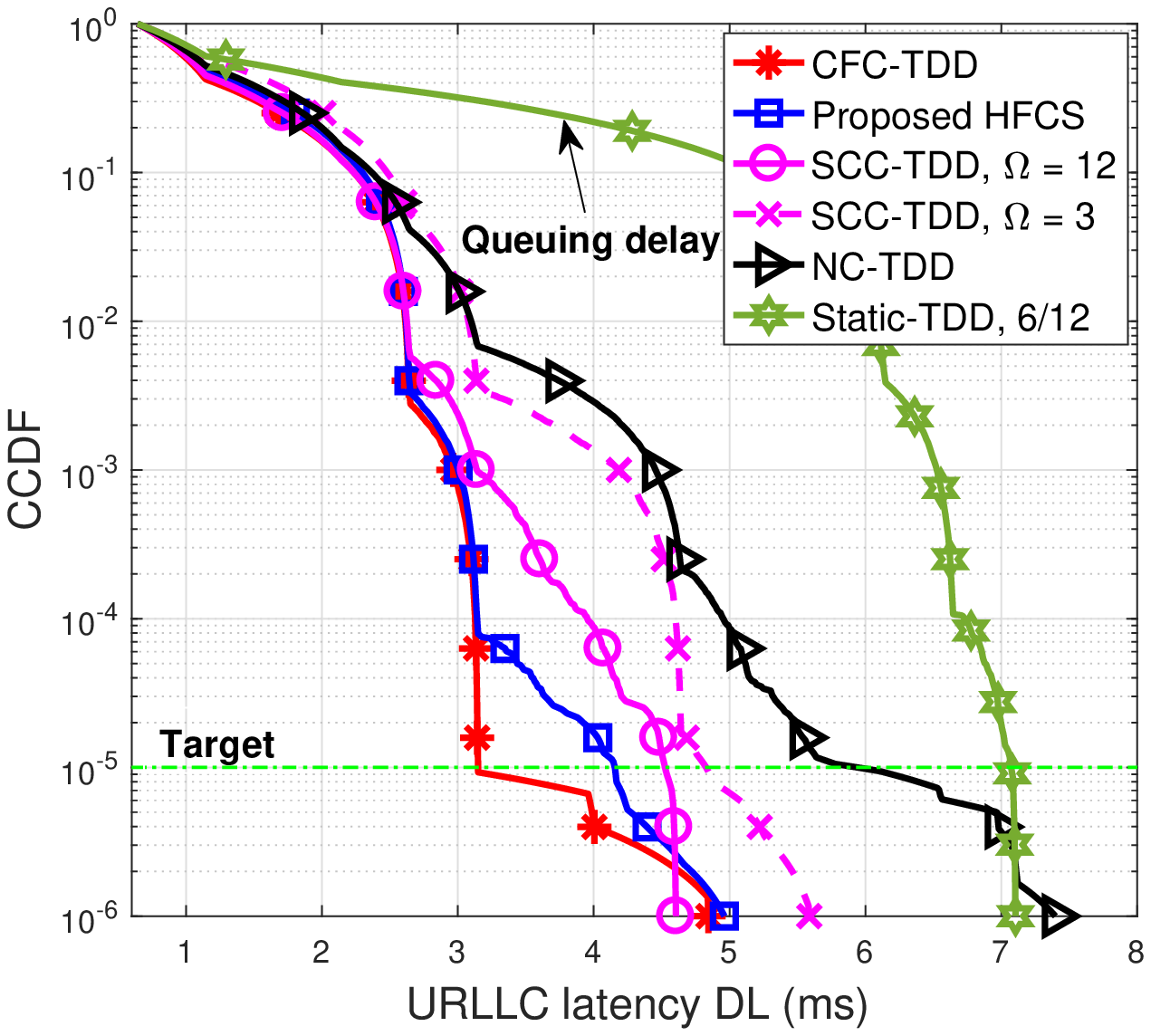}
\par\end{centering}
\centering{}{\small{}Fig. 4. URLLC outage latency in DL direction
(ms).}{\small \par}
\end{figure}

Similar observations are obtained from the URLLC outage latency in
the DL direction, as shown in Fig. 4. All considered TDD coordination
schemes provide a decent DL latency , i.e., $\leq8$ ms. This is due
to the larger desired DL transmission power, i.e., compared to the
interfering UL power, hence, less impactful CLI. However, static-TDD
case inflicts a longer queuing delay due to the fixed DL and UL slot
placement. 

Fig. 5 shows the empirical CDF (ECDF) of the post-receiver UL interference
performance in dBm, including both cross and same link inter-cell
interference, respectively. Due to the absence of the CLI, CFC-TDD
offers an attractive interference performance. However, due to the
dual-scheduling metrics during the DSS and SSS periods, the proposed
HFCS achieves the same interference suppression capability as the
optimal CFC-TDD for the critical lower percentiles below $20\%$,
i.e., cell-edge users. Furthermore, the proposed HFCS offers $39\%$
and $45\%$ reduction of the post-receiver interference at the $20^{th}$
percentile, compared to SCC-TDD and NC-TDD, respectively. The SCC-TDD
exhibits a monotonic interference suppression performance where cell-edge
users, get most impacted, while NC-TDD inflicts the worst interference
performance due to the extreme slot misalignment, hence, the sever
CLI levels. 

\begin{figure}
\begin{centering}
\includegraphics[scale=0.6]{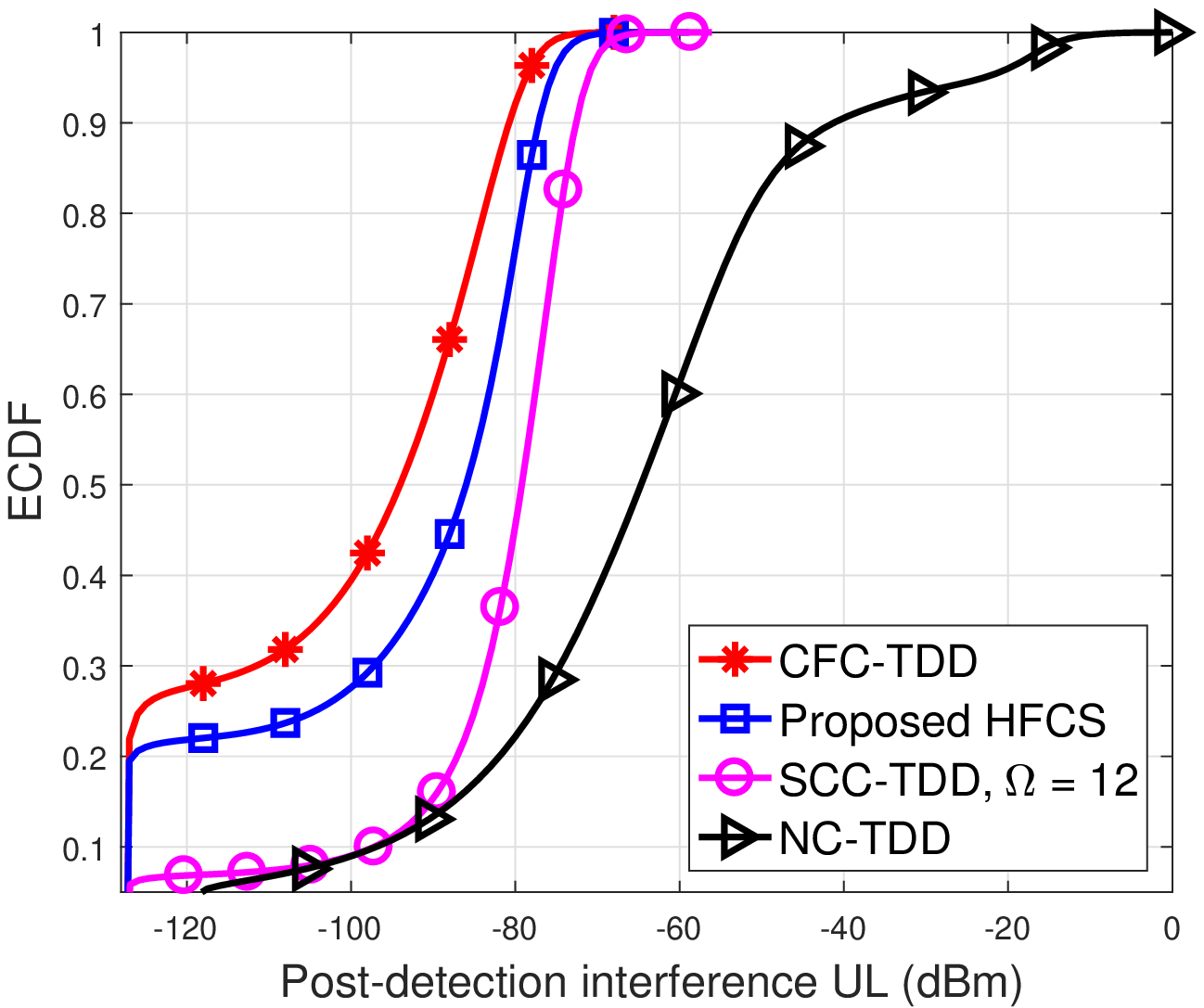}
\par\end{centering}
\centering{}{\small{}Fig. 5. Post-receiver interference in UL direction
(dBm).}{\small \par}
\end{figure}

Fig. 6 presents the average cell throughput per TTI in the UL direction,
with an average total offered load per cell of 10 Mbps. As can be
noted, proposed solution boosts the cell-edge capacity, e.g., $189\%$
capacity gain is achieved against SCC-TDD at the $30^{th}$ percentile.
The change of the distribution slope of the proposed HFCS is due to
the slot-based dual scheduling objectives, i.e., joint latency-capacity
scheduling. However, the proposed HFCS still exhibits a capacity loss
of $45\%$ at the $95^{th}$ percentile compared to ideal SCC-TDD,
due to the preemptive scheduling of cell-edge users during the SSS
of each RFC, despite that they may not be the best capacity/fairness
maximizing set of users. The fully dynamic NC-TDD fails to offer an
acceptable cell-edge capacity due to the extreme CLI, i.e., $\sim48\%$
of the scheduling TTI instances have no sufficient capacity. 

\begin{figure}
\begin{centering}
\includegraphics[scale=0.6]{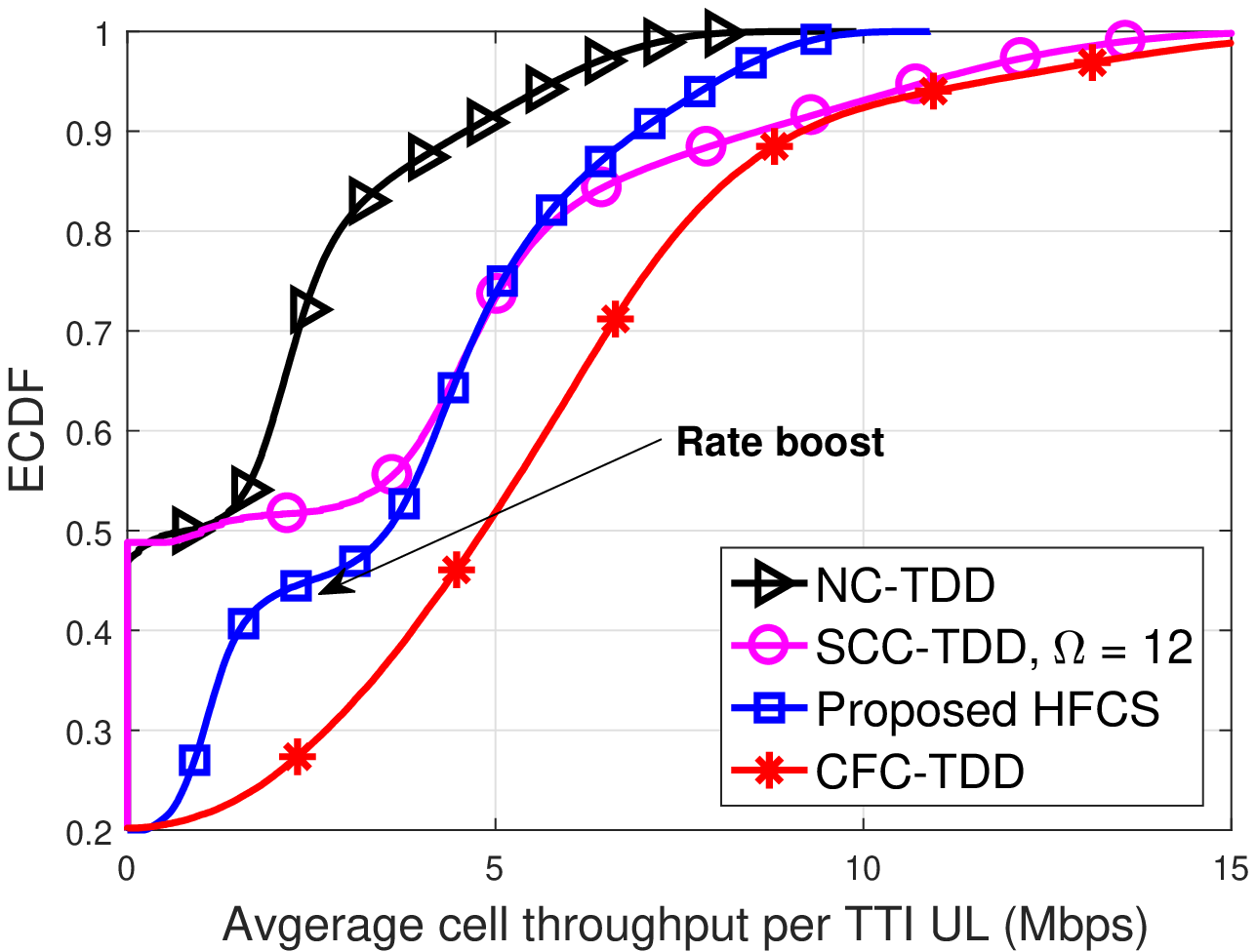}
\par\end{centering}
\centering{}{\small{}Fig. 6. Average cell throughput in UL direction
(Mbps).}{\small \par}
\end{figure}

\section{Concluding Remarks }

A quasi-dynamic coordination scheme has been introduced for ultra-reliable
and low-latency communications (URLLC) in 5G TDD networks. The proposed
solution combines hybrid radio frame design, distributed multi-objective
user scheduling and a cyclic-offset-based radio frame code-book. Compared
to the state-of-the-art coordinated TDD proposals from industry and
academia, proposed scheme offers a significant improvement of the
URLLC outage performance, e.g., $92\%$ latency reduction gain, in
addition to achieving aggregated cell capacity gain of $189\%$, and
with a limited control overhead space, bounded to $\textnormal{B-bit}.$

\section{Acknowledgments}

This work is partly funded by the Innovation Fund Denmark \textendash{}
File: 7038-00009B. Also, part of this work has been performed in the
framework of the Horizon 2020 project ONE5G (ICT-760809) receiving
funds from the European Union.

\end{document}